\documentclass[aps,prl,reprint,superscriptaddress,showpacs,showkeys,10pt,twocolumn]{revtex4-2} 
\usepackage{times}
\usepackage{amsfonts}
\usepackage{amssymb}
\usepackage{amsmath}
\usepackage{graphicx}
\usepackage{color}
\usepackage{bm}
\usepackage{xfrac}
\usepackage{array}
\usepackage{makecell}

\usepackage{multirow}
\usepackage[colorlinks=true,linkcolor=blue,urlcolor=blue,citecolor=blue]{hyperref}

\begin{document}

\title{{\color{black}Unambiguous signature of exchange interactions between nanoparticles in a collective system}}

\author{C.~A.~Iglesias}
\email[Electronic address: ]{iglesias@fisica.ufrn.br}
\affiliation{Departamento de F\'{i}sica, Universidade Federal do Rio Grande do Norte, 59078-900 Natal, RN, Brazil}
\author{J.~C.~R.~de~Ara\'{u}jo}
\affiliation{Departamento de F\'{i}sica, Universidade Federal do Rio Grande do Norte, 59078-900 Natal, RN, Brazil}
\author{J.~Xavier}
\affiliation{Departamento de F\'{i}sica, Universidade Federal do Rio Grande do Norte, 59078-900 Natal, RN, Brazil}
\author{R.~B.~da~Silva}
\affiliation{Departamento de F\'{i}sica, Universidade Federal do Rio Grande do Norte, 59078-900 Natal, RN, Brazil}
\author{J.~M.~Soares}
\affiliation{Departamento de F\'{i}sica, Universidade do Estado do Rio Grande do Norte, 59610-090 Mossor\'{o}, RN, Brazil}
\author{L.~M.~Silva}
\affiliation{Departamento de F\'{i}sica, Universidade Federal do Rio Grande do Norte, 59078-900 Natal, RN, Brazil}
\author{J.~H.~de~Ara\'{u}jo}
\affiliation{Departamento de F\'{i}sica, Universidade Federal do Rio Grande do Norte, 59078-900 Natal, RN, Brazil}
\author{S.~N.~de~Medeiros}
\affiliation{Departamento de F\'{i}sica, Universidade Federal do Rio Grande do Norte, 59078-900 Natal, RN, Brazil}
\author{P.~B.~Souza} 
\affiliation{Departamento de F\'{i}sica, Universidade Federal de Santa Catarina, 88040-900 Florian\'{o}polis, SC, Brazil}
\author{C.~C.~Pl\'{a}~Cid} 
\affiliation{Departamento de F\'{i}sica, Universidade Federal de Santa Catarina, 88040-900 Florian\'{o}polis, SC, Brazil}
\author{M.~Gamino}
\affiliation{Departamento de F\'{i}sica, Universidade Federal do Rio Grande do Norte, 59078-900 Natal, RN, Brazil}
\author{M.~A.~Correa}
\affiliation{Departamento de F\'{i}sica, Universidade Federal do Rio Grande do Norte, 59078-900 Natal, RN, Brazil}
\author{F.~Bohn}
\email[Electronic address: ]{felipebohn@fisica.ufrn.br}
\affiliation{Departamento de F\'{i}sica, Universidade Federal do Rio Grande do Norte, 59078-900 Natal, RN, Brazil}

\date{\today}

\begin{abstract}
{\color{black}
We report a simple, efficient manner to assess magnetic interactions, more specifically the exchange interaction, in systems as blocked magnetic nanoparticles. 
Specifically, we investigate what is the theoretical limit for the external magnetic susceptibility in a system described by the plain old Stoner-Wohlfarth model. 
We go beyond and introduce a general mean field theory for interacting systems, thus estimating how the magnetic susceptibility is affected due to the dipolar and exchange interactions inside the system. 
We disclose a fundamental inequality for the magnetic susceptibility and show its violation is an unambiguous signature of the existence of exchange interactions between nanoparticles in a collective system. 
To test the robustness of our theoretical achievements, we examine magnetization measurements and Henkel plots for blocked magnetic nanopowders. 
The agreement between experiment and theory provides evidence to confirm the validity of our findings. 
}
\end{abstract}

\maketitle

\section{Introduction}

For more than a century, the dependence of the magnetization with magnetic field has attracted attention as a signature of magnetic materials, bringing fundamental insights on the physical mechanisms involved in the dynamics. 
On the theoretical side, approaches have been developed to address the magnetic properties of magnets, capturing essential features of the magnetization process~\cite{Usov2009, Carrey2011, MagnodeLimaAlves2017, Verde2012}. 
One celebrated example that has shaped our thinking despite its simplicity is the Stoner-Wohlfarth (SW) model~\cite{Stoner1948}, where the magnetic behavior of non-interacting uniaxial-anisotropy particles is scrutinized. 
For interacting systems in turn, the presence of dipolar and/or exchange interactions affecting the magnetization dynamics makes the description of the whole magnetic properties a hard task. 
More than that, the identification process of the kind of interaction is itself a challenge, given the complexity of the employed routines~\cite{Geshev1992, Garcia-Otero2000, JMMM467p135, JMMM500p166420}. 

Within this framework, experimentally magnetic nanoparticles and their wide diversity provide a fascinating playground for investigating the magnetic properties in both non-interacting and interacting systems. 
Magnetic nanoparticles have drawn attention for several decades due to their challenging physical properties and potential of application~\cite{Bean1959, Guimaraes2009}. 
Perhaps most of the progress on the discovery of novel materials and the exploration of the dynamic magnetic response in diverse particle systems has been driven by the technological demand. 
And, within such perspective, magnetic nanoparticles have appeared for instance in the context of biomedical engineering~\cite{Bekovic2010a,Krishnan2010,Iglesias2021,DiCorato2014, Pankhurst2003,Kumar2011,Stimphil2017,Chung2004}, as well as in a wide variety of applications~\cite{Govan2014,Zargoosh2013,Komarek2009,Zayat2009}. 

Nevertheless, recent advances in the field of magnetization dynamics have stimulated renewed interest in phenomena involving the interactions between magnetic nanoparticles. 
For magnetic systems, it is well known their magnetic properties are dependent on numerous issues, including experimental parameters employed in the production of the sample as well as features owing to the chemical composition~\cite{DiCorato2014, Tomitaka2012, JMMM391p83, PB488p43, ML236p526, CSA560p376, Araujo2021, Xavier2022}. 
For nanoparticles, in addition, the interactions between particles have key role on the whole magnetic properties and magnetization dynamics~\cite{Fu2018, Allia2001, Sanchez2017, Botez2015, Vieira2019, Aslibeiki2010, 
Pedrosa2018, MagnodeLimaAlves2017, Chantrell1999a, Dormann1999}. 
The presence and intensity of such interactions have been often explored by employing remanence plots~\cite{Klik1997, Garcia-Otero2000, Basso1994, Soares2011, Thamm1998}. 
While several aspects of the interactions have been subject to recent analysis, it remains unclear whether or not there is a simple, straight way to relate general features of the system, as the magnetic susceptibility, and kind of interaction between particles contributing to its properties and dynamics. 

{\color{black}
In this article, we derive a simple, efficient manner to assess magnetic interactions, more specifically the exchange interaction. 
We first introduce a general mean field theory for interacting systems and estimate how the magnetic susceptibility is affected due to the dipolar and exchange interactions inside the system. 
Then, we investigate what is the theoretical limit for the external magnetic susceptibility in a Stoner-Wohlfarth system and show the violation of a fundamental inequality for susceptibility is a signature of the existence of exchange forces between nanoparticles in a collective system. 
We compare our theoretical achievements with parameters obtained experimentally for blocked magnetic nanopowders; 
and the agreement between experiment and theory directly confirm the validity of our findings. 
}

\section{Theoretical approach}

We start our approach by recalling some well-known energy-work relations involved in the magnetization process~\cite{Cullity}. 
The relations between energy and work achieved from a macroscopic point of view are of utmost importance for evaluating the energy contributions coming from the interactions between nanoparticles in a collective system. 

In order to estimate the work undergone by magnetic nanoparticles in the magnetization process, for now, let us suppose a system consisting of a magnetic sample having cylindrical form, with radius $R$ and length $l$; additionally, let us consider a solenoid with $N$ turns, radius $R$ and length $l$ too, wound around the sample. 
For sake of simplicity, we yet assume $l\gg R$, so that it may be taken as an ideal solenoid. 
Under these assumptions, the magnitude of the magnetic field $H$ inside the solenoid may be written as~\cite{David}
\begin{equation}
H = \frac{N i}{l},
\label{Hsolenoid}
\end{equation}
\noindent in which and $i$ is the electrical current passing through the solenoid. 

The solenoid is connected to an electrical current source, which provides a voltage $V$ in a time interval $dt$ and produces a current $i$ in the solenoid. 
In the magnetization process, as soon as the current $i$ is modified by an amount $di$, we observe a change of magnetic flux $ \Phi $ inside the solenoid, and consequently throughout the sample. 
Keeping in mind we still remain with ideal conditions, assuming the field is homogeneous through the cross-sectional area $ A $ of the sample, the magnetic flux may be simply written as $\Phi = A B$, where $ B $ is the magnetic induction. 
Then, the change of magnetic flux inside the solenoid may be expressed simply as 
\begin{equation}
\frac{d\Phi}{dt} = A\frac{dB}{dt}. 
\label{dPhi}
\end{equation}
Such change in the flux causes a back electromotive force $ v $ in the own solenoid, which by means of the Faraday's Law~\cite{David} is
\begin{equation}
v = -Nd\Phi/dt, 
\label{faraday}
\end{equation}
\noindent being $v = -V$, and work must be done to overcome this back electromotive force~\cite{Cullity}. 

In this framework, the work per unit volume, $ w $, done by electrical current source in a time interval $ dt $ to overcome the back electromotive force on the solenoid wire can be written in differential form as 
\begin{equation}
	dw = \dfrac{1}{V_{s}} P dt = \dfrac{1}{V_{s}} V i dt,
	\label{dw}
\end{equation}
in which $ V_{s} $ is the inner volume of the solenoid, and $ P = Vi $ is the electrical power delivered by the source to the circuit. 
Notice that, since we are interested only in the magnetic effects, we neglect in Eq.~(\ref{dw}) the electrical resistance of the solenoid and the correspondent Joule effect contribution to the energy balance. 
From Eq.~(\ref{dw}), and taking into account Eqs.~(\ref{Hsolenoid}), (\ref{dPhi}) and (\ref{faraday}), {the electrical work per unit volume done by the electrical source} can be expressed in a generalized differential form as 
\begin{equation}
dw = H\,dB,
\label{dw2}
\end{equation}
with $ B = \mu_{0} (H + m) $, where $ \mu_{0} $ is the magnetic permeability of free space and $ m $ is the volumetric magnetization of the sample. 
Without loss of generality, we use simply in Eq.~(\ref{dw2}) the magnitudes $H$ and $B$, assuming the fact the magnetization vector $ \vec{m} $ is the volumetric average of the magnetic moment of the whole sample along direction defined by the external magnetic field vector $ \vec{H} $. 

From it we can explore straightly some cases of interest.
The first consists in the case in which the solenoid is empty, without the presence of the sample. 
Here, the total electrical work per unit volume done by the electrical source is totally converted to potential energy, remaining stored in the magnetic field. 
Then, assuming $B= \mu_{0} H $ and integrating Eq.~(\ref{dw2}) from $ H=0 $ up to a given $  H $ value, we may express {the conservative energy per unit volume stored in the magnetic field $ u_{m} $} as  
\begin{equation}
u_{m} = \dfrac{1}{2} \mu_{0} H^{2}.
\label{um}
\end{equation}

Next, the second one corresponds to the case in which the sample is inside the solenoid. 
By taking into account the general Principle of Energy Conservation, we can split the total electrical work per unit volume done by the electrical source into two components, the aforementioned energy per unit volume stored in the magnetic field $ u_{m} $ and the work per unit volume undergone by the sample, $ w_{s} $. 
Hence, in differential form, $ dw_{s} = dw - du_{m} $, which becomes after integration
\begin{equation}
w_{s} = \mu_{0} \int H\, dm.
\label{ws}
\end{equation}
Such equation is the general expression for obtaining {the work undergone by a magnetic system due to its interaction with the magnetic field}. 

Going beyond, once the work undergone by a macroscopic magnetic system is established, we address here the case of a collective system of interacting magnetic nanoparticles.
To this end, let us assume the sample is small enough, relative to the detecting system of the experimental setup, to be considered a point dipole. 
Notice that such condition is often satisfied when the experiment is carried out by means of vibrating-sample or SQUID magnetometers~\cite{Cullity}. 
This assumption is a key factor that allows us to introduce in our theoretical approach the demagnetizing mean field theory proposed by Sánchez and collaborators~\cite{Sanchez2017} and the well-known Weiss's mean field theory~\cite{Cullity, Coey, Stanley1972}; as a consequence, both the dipolar and exchange interactions between magnetic nanoparticles inside the sample are inserted naturally in the model. 
From this perspective, the magnitude of the internal mean field $ H_{in} $ in the sample may be expressed as
\begin{equation}
	H_{in} = H + H_{m} + H_{d} = H + \gamma m - \gamma_{d} m.
	\label{Hin}
\end{equation}
Here $ H_{m} = \gamma m $ and $ H_{d} = -\gamma_{d} m $ are the mean fields associated with the exchange and dipolar interactions, respectively, with $ \gamma $ being the effective mean field constant and $ \gamma_{d} $ corresponding to the effective demagnetizing factor.
Given the same aforementioned reasons, we keep using the magnitudes $H$ and $m$, instead of the vectorial form. 
It is worth mentioning we understand each nanoparticle behaves as a macrospin. 
This condition implies in the coherent magnetization rotation of each nanoparticle, i.e.\ the magnetic free energy associated with the exchange forces inside each single domain is always minimized. 

Then, from Eqs.~(\ref{ws}) and~(\ref{Hin}), the work per unit volume undergone by the nanoparticles system, i.e.\ the collective system of interacting blocked magnetic
nanoparticles, may be rewritten as 
\begin{equation}
w_{np} = \mu_{0} \left[\int H \, dm + \int (\gamma - \gamma_{d}) m \, dm \right].
\label{wnp1}
\end{equation}
\noindent In the linear magnetization regime, in which $ m = \chi H $, with $ \chi $ being the {\it external} magnetic susceptibility, the integration of Eq.~(\ref{wnp1}) from $ m =0 $ to a given $ m$ value provides 
\begin{equation}
w_{np} = \dfrac{1}{2} \mu_{0}mH + \dfrac{1}{2} \mu_{0}(\gamma - \gamma_{d}) m^{2}.
\label{wnp}
\end{equation}
\noindent In this case, the first term on the r.h.s.\ of Eq.~(\ref{wnp}) represents the interaction of the system with the external magnetic field; the second term in turn brings to light the energy contributions due to magnetic interactions between nanoparticles inside the system. 

Further, we can represent the mean field constant $ \gamma $ in terms of the exchange integral $ J_{ex} $. 
For sake of simplicity and without loss of generality, we assume here the simplest case, in which only the surface atomic moments of each nanoparticle interact with the nearest neighbors on the surface of the adjacent nanoparticles. 
As a result, the mean field constant is given by $ \gamma = 2J_{ex} Z/\mu_{0} n g^{2} \mu_{B}^{2} $~\cite{Coey}, and Eq.~\eqref{wnp} may be expressed as 
\begin{equation}
w_{np} = \dfrac{1}{2} \mu_{0}mH + J_{ex} \dfrac{Z}{n g^{2} \mu_{B}^{2}}m^{2} - \dfrac{1}{2} \mu_{0} \gamma_{d} m^{2},
\label{wnp2}
\end{equation}
\noindent where $ Z $ is the number of nearest neighbors of each atomic magnetic moment, $ n $ corresponds to the number of atomic moments per unit volume, $ g $ is the g-factor and, at last, $ \mu_{B} $ is the Bohr magneton.

It is worth highlighting that here the increase of magnetization $ m $ is actually a function of the parameters $ H $, $ \gamma $ and $ \gamma_{d} $, i.e. $ m = m(H, \gamma, \gamma_{d}) $. 
Thereby, Eq.~(\ref{wnp}) corresponds to the general expression for { the work undergone by an interacting blocked nanoparticles system due to the internal mean field, taking into account the contribution of the external magnetic field combined with the ones of ferromagnetic exchange and dipolar interactions}. 

As expected, for a non-interacting magnetic nanoparticles system yet in the linear magnetization regime, Eq.~(\ref{ws}), after integration from $ m=0 $ to $ m=m_{0} $, comes to 
\begin{equation}
w_{0} = \dfrac{1}{2} \mu_{0}m_{0} H.
\label{w0}
\end{equation}
Here, we used $ m_{0} = m(H) $ to represent the increase of the volumetric magnetization of a non-interacting system to make different from the interacting case, $ m = m(H, \gamma, \gamma_{d}) $. 
Notice that for a classical Stoner-Wohlfarth-like system~\cite{Stoner1948}, for instance a non-interacting system consisting of single domain blocked particles, the work $ w_{0} $ is done only against the individual anisotropy forces that arise in each particle due to the orientation of the magnetic field in a non-parallel direction with respect to the anisotropy axes. 
In this context, we must point out Eq.~(\ref{w0}) is valid just for the conditions in which the magnetic field is below the anisotropy field $H_k$, i.e.\ $ H \ll H_{k} $.

As previously mentioned, the magnetization in the linear regime can be written in terms of the external magnetic susceptibility through $ m = \chi H $. 
Coming back to interacting systems, the {\it internal} magnetic susceptibility $ \chi_{in} $, in the linear regime, may be expressed as
\begin{equation}
\chi_{in} = \dfrac{m}{H_{in}} = \dfrac{m}{H + (\gamma - \gamma_{d}) m} .
\label{xin}
\end{equation}
Therefore, matching both we may relate $ \chi_{in} $ with $ \chi $ as
\begin{equation}
\chi_{in} = \dfrac{\chi}{1 + (\gamma - \gamma_{d}) \chi} .
\label{xin2}
\end{equation}
From Eq.~(\ref{xin}), multiplying numerator and denominator by $ \sfrac{1}{2} \mu_0 H$ and using Eq.~(\ref{xin2}), after some algebra, we may rewrite the external magnetic susceptibility as 
\begin{equation}
\chi = \dfrac{\frac{1}{2}\mu_{0}mH}{\frac{1}{2}\mu_{0}H^{2}},
\label{x}
\end{equation}
where $ m = m(H, \gamma, \gamma_{d}) $.

Therefore, only for non-interacting nanoparticles systems, in which $m (H, \gamma=0, \gamma_d=0) = m_{0}(H)$, we verify the external magnetic susceptibility may be written as a ratio of the work undergone by the system (Eq.~(\ref{w0})) to the energy stored in the magnetic field (Eq.~\eqref{um}), i.e.,
\begin{equation}
\chi = \dfrac{w_{0}}{u_{m}}.
\label{x-n-int}
\end{equation}
In this case, taking into account the general Principle of Energy Conservation, we conclude that 
\begin{equation}
\chi \leqslant 1
\label{x2}
\end{equation} 
for the magnetization curve in the linear regime measured from the sample initially exhibiting zero magnetization. 
Notice that, in order to make easier a direct comparison between theory and experiment, we need to make use of conventional SI units. 
Such inequality, in principle valid only for non-interacting magnetic systems, represents the theoretical limit of the external magnetic susceptibility in an ideal nanoparticles system described by the plain old random Stoner-Wohlfarth model~\cite{Stoner1948}. 
Hence, {\it the violation of the inequality is a necessary and sufficient condition to prove the existence of interactions between the magnetic nanoparticles in a collective system.} 
Additionally, {\it the violation of such inequality is also an evidence of the major contribution of ferromagnetic exchange interactions between blocked nanoparticles.} 
This insight comes from Eqs.~(\ref{wnp}) and (\ref{wnp2}), where the mean field constant $\gamma$ associated with the exchange interactions is the unique that may raise the value of the magnetization $ m $, as we can confirm in Eq.~(\ref{xin}). 

\section{Experimental results}

To test the robustness of our theoretical findings, we compare them with experimental parameters acquired from magnetization measurements for different magnetic nanopowders. 
Two of them are magnetic nanoparticles of magnesium ferrite ($\rm MgFe_{2}O_{4}$) and cobalt ferrite ($ \rm CoFe_{2}O_{4}$), whilst the third consists of nanorods of barium hexaferrite ($\rm  BaFe_{12}O_{19}$). 
All of them behave like classical blocked magnetic systems; and from isothermal magnetization curves and remanence plots, we explore how the nature of interactions between nanoparticles influences the magnetic behavior and whether the kind of interaction affects the magnetic susceptibility of the system. 
Further details on the samples and experiments are given in the Appendix.

Figure~\ref{Fig_01} shows the magnetization response at room temperature for our nanopowders. 
Specifically, we take special care to first obtain the magnetic response from the sample initially exhibiting zero magnetization, and then acquire the magnetization loops. 
From virgin or AC demagnetized samples, we may achieve reliable estimate of the external magnetic susceptibility, the quantity that we aim to test. 
The inset in Fig.~\ref{Fig_01} depicts the magnetization curves at low fields measured from the demagnetized samples, and brings to light how we determine the external magnetic susceptibility. 
The magnetic parameters obtained from the magnetization curves at room temperature for our $\rm MgFe_{2}O_{4}$, $ \rm CoFe_{2}O_{4}$ and $\rm  BaFe_{12}O_{19}$ nanopowders are summarized in Tab.~\ref{Tab_01}. 

We start addressing the remanent magnetization $m_r$, saturation magnetization $m_s$, and the coercive field $H_c$. 
From our concern at a first moment, an important parameter for our goals could be the $H_c/H_k$ ratio. 
From the theoretical side, Stoner-Wohlfarth model~\cite{Stoner1948} predicts $H_c/H_k =0.48$ for an assembly of particles which are non-interacting single domains with uniaxial magnetic anisotropy. 
Experimentally, an estimate of the coercive field $H_c$ is quite trivial to be taken from magnetization measurements, as we can see in Tab.~\ref{Tab_01}.
However, this is often not the case for the anisotropy field $H_k$, whose obtainment methods may be quite complex. 
Due to this experimental difficulty and the complexity of its analysis, we understand the $H_c/H_k$ ratio cannot be assumed in principle as a key parameter giving us decisive information on the interactions inside the system. 

Next, another important parameter for our goals could be the $ m_{r}/m_{s} $ ratio. 
From the magnetization loops, the saturation magnetization is achieved here through fittings with the law of approach of magnetization to saturation~\cite{Cullity}, since none of the samples saturate even with the highest magnetic field values applied in the experiment. 
Despite the absence of saturation, the remanent magnetization results are consistent; they are indeed not related to minor loops, because the magnetization curves reach at least the region of reversible magnetization rotation for all of our samples. 
\begin{figure}[!t] 
	\begin{center}
		\hspace*{-0.5cm}
		\includegraphics[width=8.5cm]{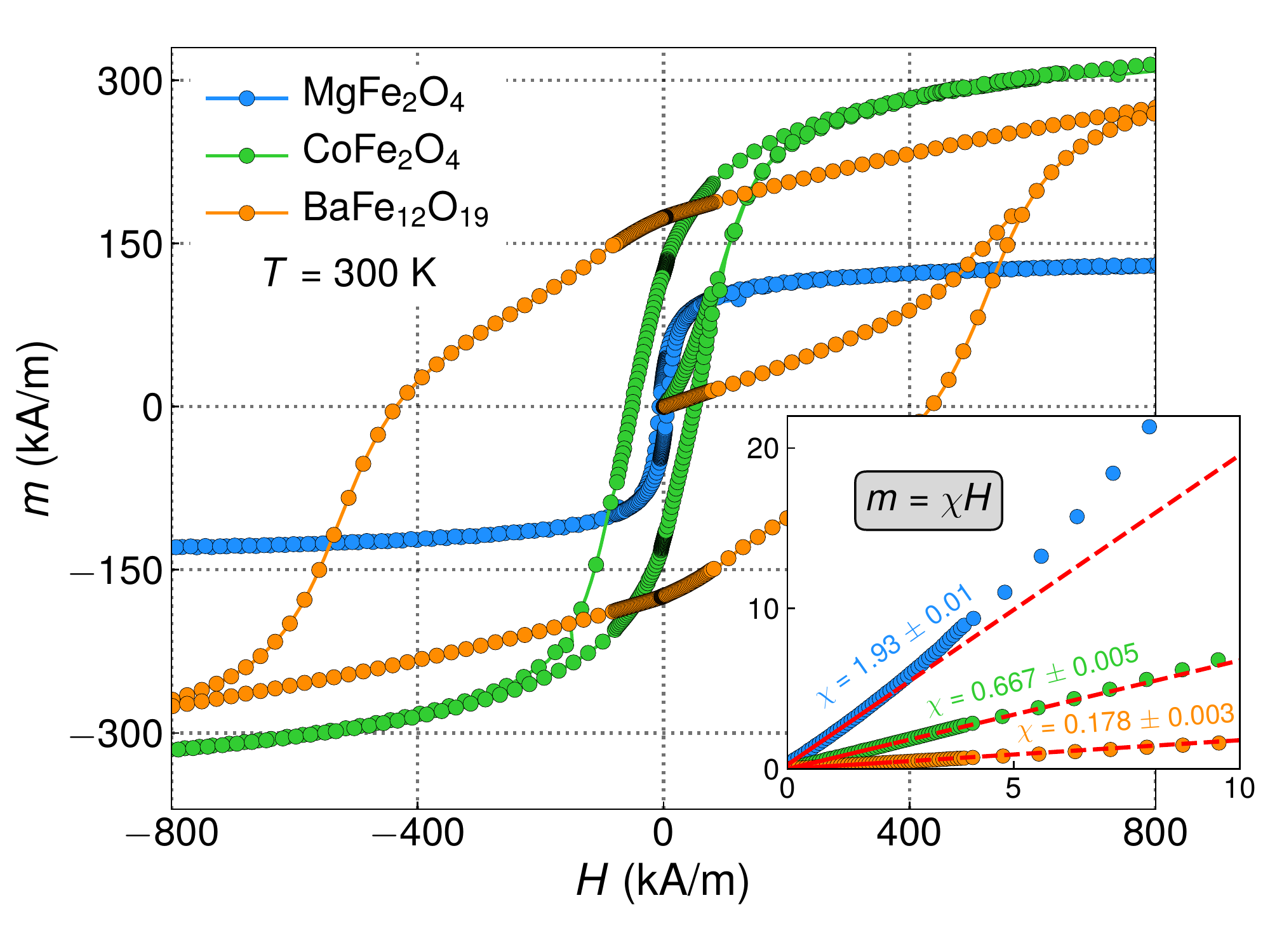}
	\end{center} 
	\vspace{-.75cm}
	\caption{Magnetization response at room temperature for the MgFe$_{2}$O$_{4}$, CoFe$_{2}$O$_{4}$ and BaFe$_{12}$O$_{19}$ nanopowders. We first obtain the magnetic response from the sample initially exhibiting zero magnetization, and then acquired the magnetization loops with maximum magnetic field of $\pm 1200$~kA/m. The inset discloses the magnetization curves at low fields measured from the demagnetized samples. Through such curves in the linear regime, we achieve an estimate of the external magnetic susceptibility. Here, the solid red lines at fields below $\sim 2.0$~kA/m correspond to the linear fits performed to determine the external magnetic susceptibility, while the red dashed lines are guides to the eyes evidencing the deviation of magnetization from the linear behavior as the magnetic field is raised. }
	\label{Fig_01}
\end{figure} 
\begin{table}[!t]\centering
	\caption{Magnetic parameters obtained from the magnetization responses at room temperature for the MgFe$_{2}$O$_{4}$, CoFe$_{2}$O$_{4}$ and BaFe$_{12}$O$_{19}$ nanopowders. Here we show the external magnetic susceptibility $ \chi $ estimated from the magnetization curve taken from the sample initially exhibiting zero magnetization, as well as the remanent magnetization $ m_{r} $, saturation magnetization $ m_{s} $, the $ m_{r}/ m_{s} $ ratio and the coercive field $H_c$ measured from the magnetization loops. }
	\begin{tabular}{ c  c  c  c  c  c c c }
		\hline \hline
		\footnotesize{\textbf{Nanopowder}} & \footnotesize{$\bm{ \chi $}} & \footnotesize{\textbf{\makecell{$ \bm{m_{r}} $\\(kA/m)}}} & \footnotesize{\textbf{\makecell{ $ \bm{m_{s}} $\\ (kA/m)}}} &
		\large{\textbf{ $\bm{\frac{m_{r}}{m_{s}}} $}} &		
		\footnotesize{\textbf{\makecell{ $\bm{H_{c}}$ \\(kA/m)}}}
		\\
		\hline 		
		\footnotesize{MgFe$_{2}$O$_{4}$} & \footnotesize{$1.93 \pm 0.01 $}  & \footnotesize{$ 31.75$} & \footnotesize{$140.6 \pm 0.3$} &
		\footnotesize{$  0.23$} &		\footnotesize{$  6.65$} \\
		\footnotesize{CoFe$_{2}$O$_{4}$} & \footnotesize{$0.677 \pm 0.005$}  & \footnotesize{$ 122.60$} & \footnotesize{$364.1 \pm 0.3$} &
		\footnotesize{$  0.34$} &
		\footnotesize{$  51.92$} \\	

		\footnotesize{BaFe$_{12}$O$_{19}$} & \footnotesize{$0.178 \pm 0.003$}  & \footnotesize{$ 164.53$} & \footnotesize{$341.4 \pm 0.8$} &
		\footnotesize{$  0.48$} &
		\footnotesize{$  442.61$} \\
		\hline \hline
	\end{tabular}
	\label{Tab_01}
\end{table} 

Stoner-Wohlfarth model~\cite{Stoner1948} yet predicts precisely $m_r/m_s = 0.5$ for an assembly of particles which are non-interacting single domains with uniaxial magnetic anisotropy. 
Curiously enough, at first glance the experimental results shown in Tab.~\ref{Tab_01} might mislead us, suggesting our BaFe$_{12}$O$_{19}$ nanorods with $m_r/m_s$ close to $0.5$ behave like a typical SW system. 
In addition, the $m_r/m_s$ values smaller than $0.5$ found for our MgFe$_2$O$_4$ and CoFe$_2$O$_4$ nanopowders would lead us to raise the hypothesis of the existence of demagnetizing effects due to dipolar interactions between the particles. 
However, such assumption does not think about contributions coming from for instance thermal effects, as well as from the magnetization mechanisms as the non-coherent rotation of the single domains. 
They are not fully discarded in our samples and in principle may promote the reduction of the $m_r/m_s$ ratio. 
In the condition in which the magnetization process does not proceed exclusively through coherent magnetization rotation as in a truly SW system, we observe deviations from the premise that the free energy associated with exchange forces inside the nanoparticles is always minimized. 
If it is the case, hence dipolar and exchange interactions between the atomic magnetic moments within the nanoparticles shall be taken into account in Eq.~(\ref{wnp2}). 
Within this scenario, whatever is the case, $ m_{r}/m_{s} $ ratio cannot be also taken as a key parameter providing us reliable information on the interactions inside the system. 

Given all the stated above, it is quite interesting that $m_r/m_s = 0.5$ and $H_c/H_k =0.48$ are often taken as references for the behavior described by the SW model, while the magnetic susceptibility is commonly left aside in the scrutiny of, for instance, nanoparticle systems. 

Then, we focus on the magnetization curves measured from the sample initially exhibiting zero magnetization. 
From linear fits performed at low fields, in which such regime is valid, we determine the external magnetic susceptibility for our nanopowders. 
From Fig.~\ref{Fig_01} and Tab.~\ref{Tab_01}, we notice two of our samples satisfy Eq.~(\ref{x2}), whereas one of them has $\chi$ above $1$. 
For both CoFe$_2$O$_4$ and BaFe$_{12}$O$_{19}$ nanopowders having $\chi$ of $0.677 \pm 0.005$ and $0.178 \pm 0.003$, respectively, the evaluation of the external magnetic susceptibility through the requirement of Eq.~(\ref{x2}) does not permit us the strict identification of the existence of magnetic interactions between particles as well as of the kind of interactions in the system. 
As a consequence, the discrimination between interacting and non-interacting systems and the assessment of the interactions shall be done following another way. 

The most striking finding here resides in the external magnetic susceptibility for our MgFe$_2$O$_4$ nanoparticles. 
Specifically, we estimate $1.93 \pm 0.01$, thus violating the inequality in Eq.~(\ref{x2}). 
It is a necessary and sufficient condition to infer existence of interactions between the magnetic nanoparticles in a collective system. 
In addition, it corresponds to an unambiguous signature of the existence of ferromagnetic exchange forces between nanoparticles. 

The question then arises how good our sample is; whether it is a single domain system in which the coherent magnetization rotation of each nanoparticle is indeed the major magnetization mechanism. 
In the case framework such is literally true, this sample behaves as a matter of fact as an interacting SW-like system, with short-range exchange interactions being the dominant sort of interaction between the nanoparticles. 
We understand it is the case, based on the morphology, particle size, and magnetic response that are typical of blocked single domain systems. 
Nevertheless, one could also raise another possibility of exchange forces acting inside the own nanoparticles, due to non-coherent rotations~\cite{Guimaraes2009, Coey}. 
Anyway, irrespective of the actual reason, which does depend for instance on the critical sizes of particles for coherent rotation and/or pseudo single domain behavior~\cite{Coey, Guimaraes2009, Hergt_2008}, as well as on the structure of the particles (for instance, single domain and core-shell), it is unquestionable the presence of ferromagnetic exchange interactions in the system. 

In order to achieve further experimental evidence to corroborate our assumptions on the dominant interaction in our systems, we proceed with the traditional $ m_{d} \,\, vs.\ m_{ir} $ curve, a.k.a.\ Henkel plot~\cite{Garcia-Otero2000,Basso1994}. 
The curve is obtained by means of isothermal remanent magnetizations (IRM) and DC demagnetization (DCD) protocols~\cite{Garcia-Otero2000,Soares2011}. 
In the IRM procedure, $ m_{ir}(H) $ is determined by measuring the remanent magnetization $m_{ir}$ of a sample initially exhibiting zero magnetization by applying and subsequently turning off a positive magnetic field $H$. 
Such procedure is repeated, and increasing magnetic fields are successively applied until the remanent magnetization reaches the saturation remanent state $ m_{ir}(\infty) $. 
In the DCD protocol, the procedure is quite similar, except for its onset being the positive saturation state. 
After the positive saturation, the quantity $ m_{d}(H) $ is here estimated by obtaining the remanent demagnetizing magnetization $ m_{d}$ by applying and subsequently turning off a negative magnetic field $H$. 
Again, such process is repeated, and increasing magnetic fields are successively applied until remanent magnetization reaches negative saturation. 
For all, we take special care and perform the procedure at low temperatures, thus avoiding undesirable thermal demagnetizing effects in the system. 

Most of studies on interactions are based on the Wohlfarth relation~\cite{Garcia-Otero2000,Soares2011,Wohlfarth1958,Basso1994,Thamm1998,Klik1997}, 
\begin{equation}
 m_{d}(H) = m_{ir}(\infty) - 2m_{ir}(H), 
 \end{equation}
 \noindent which can be rewritten as
 \begin{equation}
 m^{,}_{d}(H) = 1 - 2m^{,}_{ir}(H), 
 \end{equation}
where $ m^{,}_{d}(H) = m_{d}(H)/m_{ir}(\infty)  $ and $ m^{,}_{ir}(H) =  m_{ir}(H)/m_{ir}(\infty) $ are the normalized remanent magnetizations.
The Wohlfarth relation should be valid for non-interacting uniaxial-anisotropy particles described by the SW model; and deviations from the Wohlfarth relation are ascribed to magnetic interactions. 
Specifically, positive deviations are attributed to exchange-like interactions, whilst negative ones are believed to indicate dipolar-like interactions~\cite{Garcia-Otero2000}. 

\begin{figure}[!t] 
 	\begin{center}
 		\hspace*{-0.5cm}
 		\includegraphics[width=8.5cm]{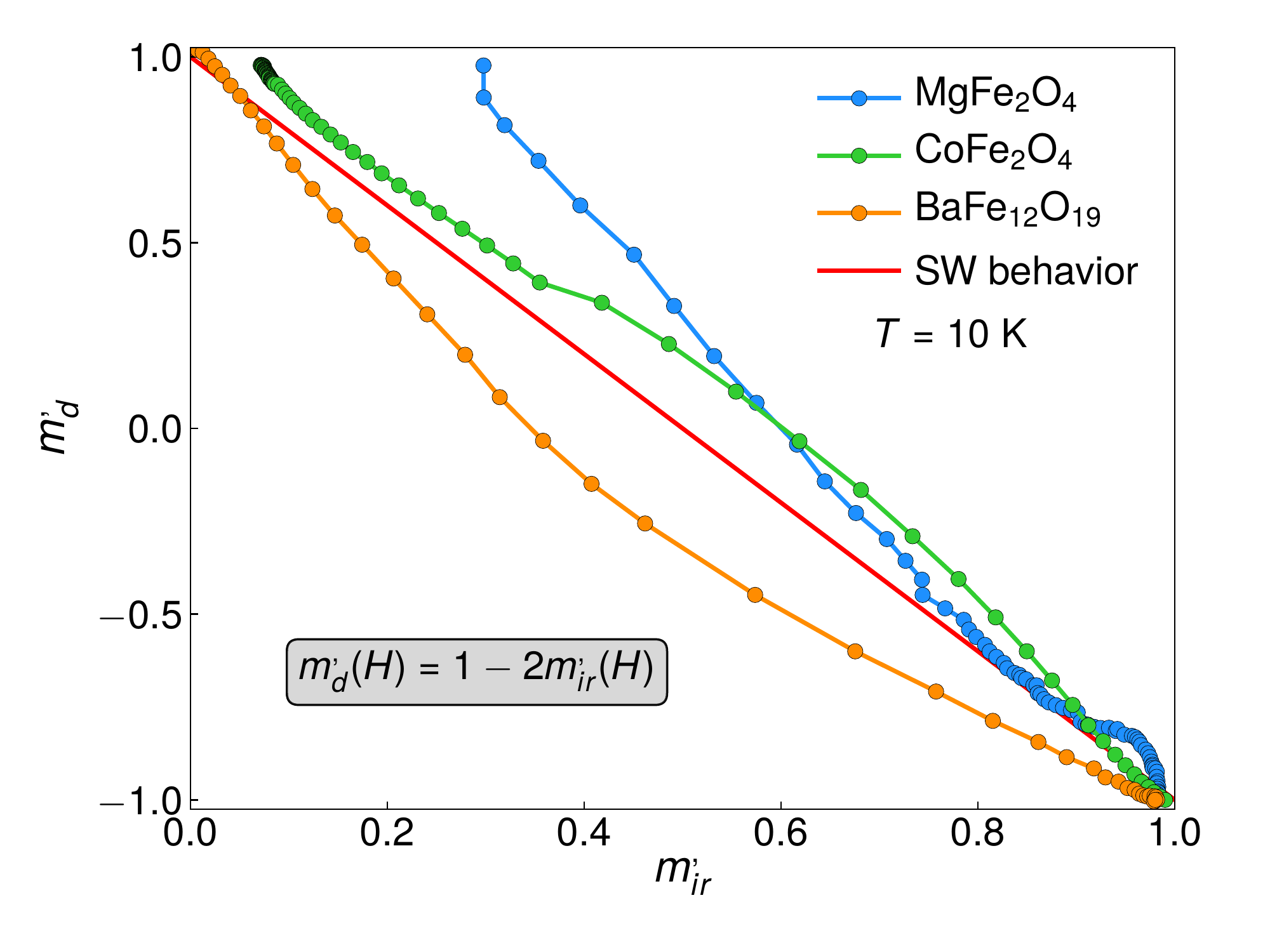}
 	\end{center} 
 	\vspace{-.75cm}
 	\caption{Henkel plot, obtained at $10$~K, for the MgFe$_{2}$O$_{4}$, CoFe$_{2}$O$_{4}$ and BaFe$_{12}$O$_{19}$ nanopowders. The red line brings the expected behavior for a Stoner-Wohlfarth system. }
 	\label{Fig_02}
 \end{figure}

Figure~\ref{Fig_02} presents the Henkel plots obtained for our nanopowders. 
For a non-interacting SW system, the expected behavior is depicted by a straight line with a slope of $-2$. 
As a consequence, the Henkel plots bring to light the existence and the kind of magnetic interactions in our samples. 

For our BaFe$_{12}$O$_{19}$ nanopowder, such analysis discloses a negative deviation from the linear SW behavior, inferring the presence of dipolar interactions between the nanorods, what is not a surprise given that $\chi <1$ for this system. 

Next, for our CoFe$_2$O$_4$, a system with $\chi <1$, it is interesting to notice that the $ m^{,}_{d} \,\, vs.\ m^{,}_{ir} $ curve exhibits a positive deviation from the linear SW behavior, suggesting the presence of exchange-like interactions between nanoparticles in the system. 
We must point out that such condition is not a contradiction with respect to our previous findings. 
Although the violation of the inequality is a necessary and sufficient condition to prove the existence of interactions between the magnetic nanoparticles, the non-violation does not lead to a conclusive answer on the interactions in a collective system having $\chi <1$. 

Last but not least, for our MgFe$_2$O$_4$ nanoparticles having $\chi >1$, the Henkel plot also exhibits a positive deviation from the linear SW behavior. 
This feature infers the magnetizing trend for $ \rm MgFe_{2}O_{4} $ sample, what is indeed an evidence of the predominance of ferromagnetic exchange interactions in the system. 
This result is consistent with our previous finding; it corroborates the violation of the inequality in Eq.~(\ref{x2}) as an unambiguous signature of the existence of exchange interactions between the nanoparticles in a collective system. 

\section{Conclusion}

In conclusion, we have derived here a simple yet efficient manner to assess magnetic interactions, more specifically the exchange interaction.  
First, we have introduced a general mean field theory for interacting systems and estimated how the magnetic susceptibility is affected due to the dipolar and exchange interactions inside the system. 
Then, we have uncovered a fundamental inequality, given by Eq. (17), which represents the theoretical limit for the external magnetic susceptibility in a Stoner-Wohlfarth system. 
To test the robustness of our theoretical achievements, we have performed magnetization experiments as well as obtained Henkel plots for $\rm MgFe_{2}O_{4}$, $ \rm CoFe_{2}O_{4}$ and $\rm  BaFe_{12}O_{19}$ nanopowders, all of them behaving like classical blocked magnetic systems. 
From samples initially exhibiting zero magnetization, we have achieved reliable estimate of the external magnetic susceptibility, the quantity that we aim to test using Eq.~(\ref{x2}).
In addition, we have confirmed through Henkel plots the nature of the interactions governing the magnetic dynamics in each sample. 
Hence, the agreement between experiment and theory provides evidence to confirm the validity of our findings. 
But of course our results on the fundamental inequality for susceptibility does not aim to replace the traditional Henkel plot in the analysis of magnetic interactions. 
Nevertheless, we have demonstrated it consists of a pre-test sharp tool.
In the case the fundamental inequality for susceptibility is violated, no further experiments are needed. 
Its violation is a necessary and sufficient condition to prove the existence of interactions, as well as it corresponds to an unambiguous signature of the existence of exchange interactions between nanoparticles in a collective system. 

\begin{acknowledgments}
The research is supported by the Brazilian agencies CNPq and CAPES
\end{acknowledgments}

\appendix

\vspace*{1cm}
\centerline{\bf Appendix}

\vspace*{.35cm}
In this Appendix, we provide information on the investigated samples and details on the nanopowders preparation; characterize the samples, and discuss structural properties of the nanopowders; and provide additional information on the experiments performed for the magnetic characterization.




\vspace{-.35cm}
\section{Nanopowders and preparation of the samples} 

\vspace{-.25cm}
We employ magnesium ferrite (MgFe$_{2}$O$_{4}$), cobalt ferrite (CoFe$_{2}$O$_{4}$) and barium hexaferrite (BaFe$_{12}$O$_{19}$) nanopowders to perform the experiments probing the magnetic properties and to test our theoretical findings. 
Specifically, the first two are magnetic nanoparticles of MgFe$_{2}$O$_{4}$ and CoFe$_{2}$O$_{4}$, whilst the third consists of nanorods of BaFe$_{12}$O$_{19}$.  

The MgFe$_{2}$O$_{4}$ nanoparticles are synthesized using the sol-gel method~\cite{Benvenutti2009, Hiratsuka1995}, with the same experimental parameters employed in the preparation of the powders investigated in Ref.~\cite{Araujo2021}, followed by heat treatment at $973$~K. 
The CoFe$_{2}$O$_{4}$ nanoparticles in turn are produced by coprecipitation method~\cite{Lee2004, Witte2016} following the procedures described in Ref.~\cite{Xavier2022}, with subsequent heat treatment at $1023$ K.
At last, the nanorods of BaFe$_{12}$O$_{19}$ are prepared through ionic coordination reaction method under the same conditions as the samples studied in Ref.~\cite{Soares2007}, followed by heat treatment at $1173$~K. 

\vspace{-.35cm}
\section{Structural properties of the nanoparticle}


\vspace{-.25cm}
The structural and morphological features of the samples are acquired by X-ray diffractometry (XRD), transmission electron microscopy (TEM) and scanning electron microscopy (SEM).
The XRD measurements are performed with a Rigaku MineFlex II diffractometer, with CuK$_{\alpha} $ radiation ($\lambda = 1.54$~\AA), in the Bragg-Brentano ($ \theta -2 \theta $) geometry. 
The diffractograms are analyzed considering the ICSD database and refined using the Rietveld method with the MAUD software~\cite{Rietveld1969}, thus allowing the identification of the phase, and providing lattice parameters, average particle size and densities. 
For our MgFe$_{2}$O$_{4}$, CoFe$_{2}$O$_{4}$ nanopowders, the phase, particle shape and the distribution of the average particle diameter are studied by TEM, obtained with a JEM-1011 microscope. 
For our BaFe$_{12}$O$_{19}$ nanopowder in turn, the morphology is explored by SEM, performed using a Zeiss Auriga 60 FEG scanning electron microscope.

\begin{figure*}[!t] 
	\begin{center}
		\includegraphics[width=8.5cm]{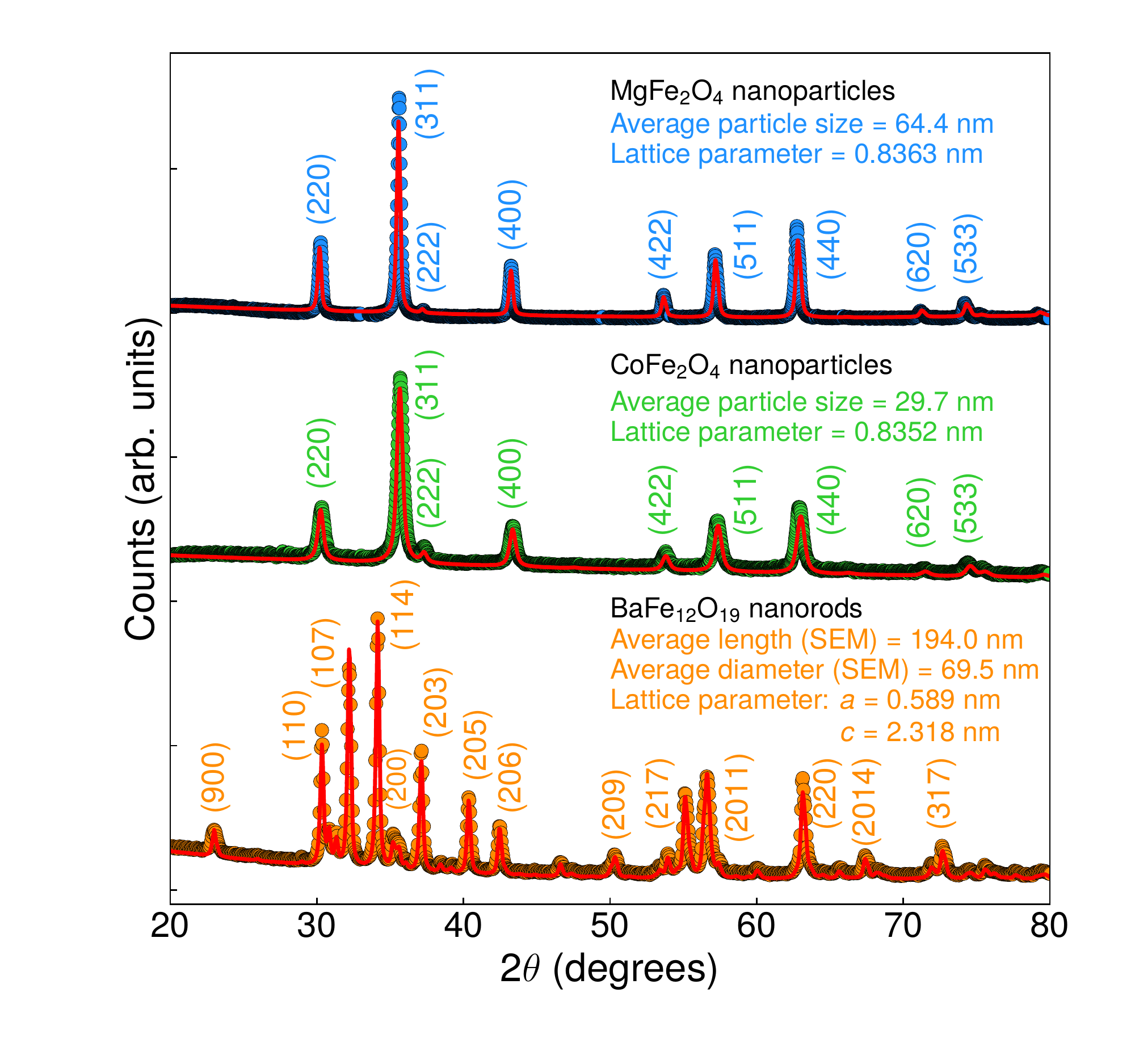}
	\end{center} 
	\vspace{-1cm}
	\caption{X-ray diffraction results for our MgFe$_{2}$O$_{4}$, CoFe$_{2}$O$_{4}$ and BaFe$_{12}$O$_{19}$ nanopowders. The symbols correspond to the experiments, and the red lines are the fits obtained with Rietveld refinement. The peaks associated to the MgFe$_{2}$O$_{4}$, CoFe$_{2}$O$_{4}$ and BaFe$_{12}$O$_{19}$ phases are indexed considering ICSD-152468, ICSD-109045 and ICSD-60984, respectively. The average particle size and lattice parameters are estimated from the Rietveld refinement; for the BaFe$_{12}$O$_{19}$ sample, the length and diameter of the nanorods are obtained by SEM. }
	\label{drx}
	\begin{center}
		\includegraphics[width=16.5cm]{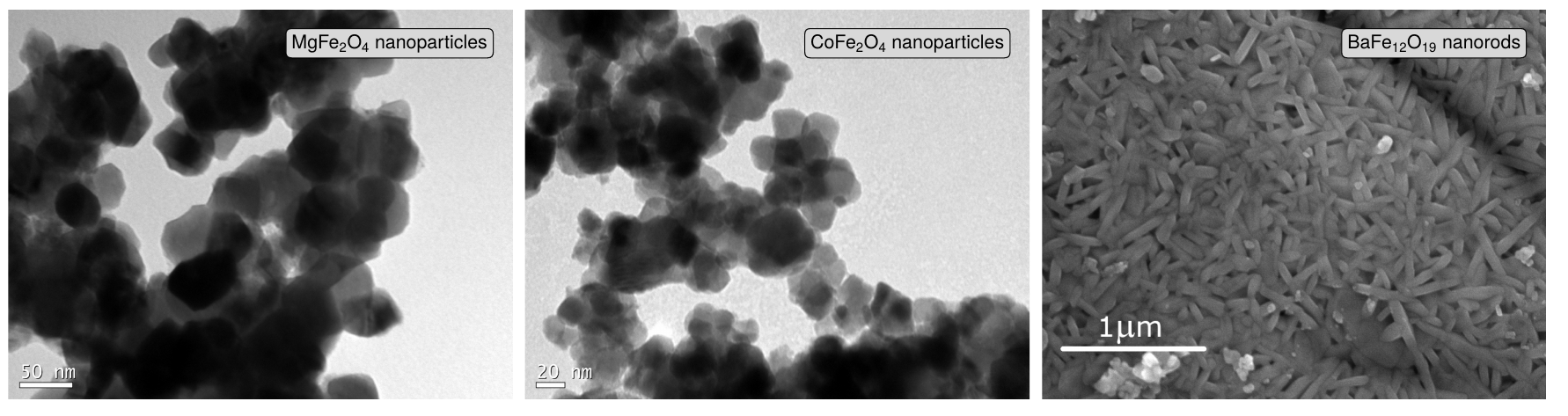}
	\end{center} 
	\vspace{-.75cm}
	\caption{Transmission electron microscopy images for our MgFe$_{2}$O$_{4}$ and CoFe$_{2}$O$_{4}$ nanoparticles, and scanning electron microscopy image for our BaFe$_{12}$O$_{19}$ nanorods. }
	\label{TEMSEM}
\end{figure*}

Figure~\ref{drx} shows the XRD results for our MgFe$_{2}$O$_{4}$, CoFe$_{2}$O$_{4}$ and BaFe$_{12}$O$_{19}$ nanopowders, whereas Fig.~\ref{TEMSEM} presents representative images of our samples. 

From Fig.~\ref{drx}, the peaks related to MgFe$_{2}$O$_{4}$, CoFe$_{2}$O$_{4}$ and BaFe$_{12}$O$_{19}$ are indexed considering the standard patterns reported in ICSD-152468, ICSD-109045 and ICSD-60984, respectively. 
From a general perspective, the results for our samples show peaks located at $2\theta$ ranging from $20^\circ$ to $80^\circ$ that are in very good concordance with the planes associated with the corresponding expected phase. 
Remarkably, the agreement between experiment and refinement confirms the formation of pure phases, without the presence of second phases or impurities.

From the refinement of the diffractograms obtained using the Rietveld method, we confirm the cubic phase of both $\rm MgFe_{2}O_{4} $ and CoFe$_{2}$O$_{4}$, with the Fd3m space group for all samples. 
We estimate average particle size of $64.6$~nm and lattice parameter of $0.8363$~nm for the MgFe$_{2}$O$_{4}$ nanoparticles; and for the CoFe$_{2}$O$_{4}$ we observe corresponding values of $29.7$~nm and $0.8352$~nm. 
Our findings are in concordance with results previously reported in literature for MgFe$_{2}$O$_{4}$~\cite{Pradeep2008,Thankachan2013,JansiRani2018,Singh2018} and 
CoFe$_{2}$O$_{4}$~\cite{Vieira2019,Verde2012} nanoparticles. 
From Fig.~\ref{TEMSEM}, 
the TEM images for $\rm MgFe_{2}O_{4} $ and CoFe$_{2}$O$_{4}$ nanopowders reveal particles with regular geometry, as well as we observe the existence of some regions with agglomerated particles.
From the TEM results, we also corroborate our findings of structural phase and average particle size obtained through XRD.

For the BaFe$_{12}$O$_{19}$ nanopowder in turn, the diffractogram and refinement shown in Fig.~\ref{drx} bring to light evidences of a M type barium hexaferrite phase belonging to P$6_{3}$/mmc space group.
In this case, we estimate lattice parameters $ a = b = 0.5895 $~nm and $ c =  2.3179 $~nm. 
From the SEM image presented in Fig.~\ref{TEMSEM} we verify this nanopowder consists of BaFe$_{12}$O$_{19}$ nanorods, having average values of length and diameter of $194.0$~nm and $69.5$~nm, respectively. 

\vspace{-.35cm}
\section{Magnetic characterization}

\vspace{-.25cm}
In the magnetic characterization, first and foremost we acquire magnetization curves at room temperature using a Lakeshore model $7404$ vibrating sample magnetometer (VSM), with maximum magnetic field of $\pm 1200$~kA/m. 
All the measurements are performed following the very same procedures. 
The system is previously calibrated using a pure Ni pattern and, after measuring the weight of each sample, provides the magnetization in A/m.
In addition, we take special care to obtain the magnetic response from the sample initially exhibiting zero magnetization. 
Such procedure allows us to achieve reliable estimate of the external magnetic susceptibility of the magnetization curve in the linear regime measured from the demagnetized sample, the quantity that we aim to test. 

Next, we use a homemade cryogenic vibrating sample magnetometer and obtain isothermal remanent magnetizations and DC demagnetization curves~\cite{Garcia-Otero2000,Soares2011} at the temperature of $10$~K. 
The IRM and DCD procedures are described in detail in the main text of the article. 
From such experiments, we achieve the traditional Henkel plot~\cite{Garcia-Otero2000,Basso1994} for our samples. 

\end{document}